# Melodic Contour and Mid-Level Global Features Applied to the Analysis of Flamenco *Cantes*

Francisco Gómez[1], Joaquín Mora[2], Emilia Gómez[3,4], José Miguel Díaz-Báñez[5]

[1]Applied Mathematics Department, School of Computer Science, Polytechnic University of Madrid, Spain

[2]Department of Evolutive and Educational Psychology, University of Seville, Spain

[3]Music Technology Group, Universitat Pompeu Fabra, Barcelona, Spain

[4]Department of Sonology, Escola Superior de Música de Catalunya, Barcelona, Spain

[5] Department of Applied Mathematics, School of Engineering, University of Seville, Spain

**Abstract**

This work focuses on the topic of melodic characterization and similarity in a specific musical repertoire: a cappella flamenco singing, more specifically in *debla* and *martinete* styles. We propose the combination of manual and automatic description. First, we use a state-of-the-art automatic transcription method to account for general melodic similarity from music recordings. Second, we define a specific set of representative mid-level melodic features, which are manually labeled by flamenco experts. Both approaches are then contrasted and combined into a global similarity measure. This similarity measure is assessed by inspecting the clusters obtained through phylogenetic algorithms algorithms and by relating similarity to categorization in terms of style. Finally, we discuss the advantage of combining automatic and expert annotations as well as the need to include repertoire-specific descriptions for meaningful melodic characterization in traditional music collections.

# 1 Introduction

## 1.1 Context and motivation

Shortly after the first MIR conference was held in 2000, in an insightful paper, Byrd and Crawford (2002) acknowledged that, "the field of MIR is still very immature." These authors advised readers of several crucial issues to be tackled for the field to blossom. Among others, they challenged the then-common assumption that searching on pitch –or pitch-contour– alone would be satisfactory enough for most purposes; they also noticed that nearly all MIR research was carried out on mainstream Western music; and they advocated the incorporation of music cognition knowledge to MIR models and techniques. Curiously enough, although Byrd and Crawford propounded models including "the four basic parameters" –namely, pitch, duration, loudness and timbre–, and even tackling subtler concepts such as salience (page 256), they failed to conceive broader theoretical frameworks comprising higher level music features –those related to harmony, voice leading, phrasing, and form. The time was not ripe. Ten years later, in a 2010 paper, Cornelis et al. (2010) examined the problem of accessing ethnic music in its full generality. In spite of all the substantial amount of research undertaken over the past few years, much of the criticism of Byrd and Crawford remains valid. MIR researchers are still focusing on Western music as Cornelis and co-authors have proved just by counting the number of papers on ethnic music presented in the 2000-2008 MIR conferences: a sad 5.5%, or 38 papers out of 686. Notice that in their study ethnic music is used in a very broad sense (music of oral tradition, non-Western classical music, and folk music). Many current models still rest upon a few music parameters and often they are combined in a flimsy, ad hoc manner. Cornelis et al. called to mind the words of Tzanetakis et al. (2007): "There is a need for collaboration between ethnomusicologists and technicians to create an interdisciplinary research field." The lack of such collaboration between researchers from both fields may be the main reason for the absence of truly music-rooted models. Also presented in Cornelis et al.'s work is a taxonomy of musical descriptors for content-based MIR (pages 1010-1011). That taxonomy comprises three broad categories: low-level descriptors, chiefly related to properties of the audio signal such as frequency, spectrum, intensity, etc.; mid-level descriptors, pertained to pitch, melody, chords, timbre, beat, meter, rhythmic patterns, etc.; and finally high-level descriptors, typically associated with meaning and expressiveness such as mood, motor and affective responses, etc. Furthermore, musical descriptors can be defined according to their scope; local descriptors refer to changes taking place in a small time span, such as note-to-note change, while global descriptors account for changes occurring on a larger scale, such as phrase division. The former are related to local features as opposed to the latter, concerned with global features. Most models do not combine descriptors taken from different levels, or trace descriptors at several scales. This results in an incomplete portrayal of musical complexity.

In this paper we have tried to meet –to a certain extent– the above criticism. This paper is concerned with flamenco music, which is music from an oral tradition; results are the fruit of collaboration between flamenco experts and technicians; our model combined generic and specific melodic descriptors for the musical repertoire under study; moreover, we integrated local and global musical descriptors, namely, melodic contour and mid-level global descriptors.

### 1.2 Goals and structure of the paper

The main goal of this paper is to study melodic characterization in flamenco music, more precisely, a cappella singing styles. Our research hypothesis is that each flamenco style is characterized by a certain prototypical melody, which can be subject to a great range of ornamentation and variation. This work researches into the most adequate way to characterize melody in this specific repertoire and analyzes the link between melodic similarity and style characterization.

Our study includes the combination of two types of descriptors for melodic characterization and similarity; it can be divided up into the following steps. First, we considered a set of mid-level musical features, specific to the repertoire, defined and manually labeled by flamenco experts. Second, we looked at melodic contour, a generic melodic descriptor, which was computed by using an automatic transcription algorithm for music recordings. Next, we compare descriptor values by using a standard similarity measure and integrate both approaches to quantify the distance between performances. Finally, we assess the obtained distances on a music collection of recordings from the most representative performers of the styles under study.

This paper is organized as follows. In the next section we analyze the characteristics of flamenco singing and a cappella singing styles. After this analysis, we address the problem of musical transcription in flamenco music and review the styles to be analyzed, providing a description of the music collection. In the fourth section we examine the problem of melodic similarity in flamenco music. Here two approaches are looked into, the musicological one and the computational one, which are presented in a top-down manner from the former to the latter. The next section contains the main contributions of the paper: the musical features of the analyzed styles are presented, the distance based on those features is described, and the combined distance is finally defined. Assessment strategies for the obtained similarity distance are thoroughly discussed and phylogenetic trees are used to visualize clustering and analyze style discrimination. A conclusion section summarizes our main findings and contributions.

## 2 Flamenco singing

### 2.1 A brief introduction.

Flamenco is an eminently individual yet highly structured form of music. Improvisation and spontaneity play a central role, but both heavily lean on an extremely stable organization of the musical material. Flamenco music has developed by coalescence of several music traditions into a rich melting pot, whose combination of singing, dancing and guitar playing is distinctive. Apart from the influences of the Jews and Arabs, flamenco music shows the imprint of the culture of the Andalusian Gypsies, who decisively contributed to its form today. We refer the reader to the books of Blas Vega and Ríos Ruiz (1988), Navarro and Ropero (1995), and Gamboa (2005) for a comprehensive study of styles, musical forms and history of flamenco.

Flamenco music was germinated and nourished mainly from the singing tradition (Gamboa, 2005). Accordingly, the singer's role soon became dominant and fundamental. In the flamenco jargon, singing is called *cante*, and songs are termed *cantes*; in this paper we use this terminology. Next, we describe the main general features of flamenco *cante*.

### 2.2 General features

Several features are characteristic of flamenco singing:

- **Instability of pitch**. In general, notes are not clearly attacked. Pitch glides or *portamenti* are very common.

- **Sudden changes in volume (loudness).** Those sudden changes are very often used as an expressive resource.

- **Short melodic pitch range.** It is normally limited to an octave and characterized by the insistence on a note and those contiguous to it.

- **Intelligibility of voices.** Lyrics are important in flamenco, and intelligibility is then desirable. For that reason, contralto, tenor, and baritone are the preferred voice *tessituras*.

- **Timbre**. Timbre characteristics of flamenco singers depend on the particular singers. As relevant timbre aspects, we can mention breathiness in the voice and absence of high frequency (singer) formants.

These characteristics contrast with classical singing styles, where precise tuning and timing are important, and where timbre is characterized by stability, absence of breathiness, and high-frequency formants (i.e., the singer formant); see Sundberg (1987).

### 2.3 Flamenco a cappella *cantes*

A cappella *cantes* constitute an important group of styles in flamenco music. They are songs without instrumentation, or in some cases with some percussion. Examples of a cappella styles are *tonás, deblas, martinetes, carceleras, nanas, saetas*, and some labor songs. Most flamenco textbooks (Molina and Mairena, 1963; Blas Vega and Ríos Ruiz, 1988) make a division between the group of *tonás* (including *tonás*, *deblas, martinetes, carceleras*) and the rest of a cappella *cantes*, which are closer to Spanish folklore (Castro Buendía, 2010).

From a musical point of view, a cappella *cantes* retain the following properties.

- **Conjunct degrees.** Melodic movement mostly occurs by conjunct degrees.

- **Scales.** Certain scales such as the Phrygian and Ionian mode are predominant. In the case of the Phrygian mode, chromatic rising of the third and seventh degrees is frequent.

- **Ornamentation**. There is also a high degree of complex ornamentation, melismas being one of the most significant devices of expressivity.

- **Microtonality.** Use of intervals smaller than the equal-tempered semitones of Western classical music.

These features are not exclusive to a cappella cantes and can be found to various degrees in other flamenco styles.

The classification of flamenco *cantes* in general and of a cappella *cantes* in particular is subject to many difficulties, and such a classification is not yet clearly established in the flamenco literature; as a case in point, compare the classifications proposed by Molina and Mairena (1963), Blas Vega and Ríos Ruiz (1988), and Gamboa (1995). Two *cantes* belonging to the same style may sound very different to an unaccustomed ear. In general, underlying each *cante* there is a melodic skeleton. Donnier (1997) called it the "*cante's* melodic gene". This melodic skeleton is filled in by the singer by using different kinds of melismas, ornamentation and other expressive resources. An aficionado's ears recognize the wheat from the chaff when listening, and appreciate a particular performance in terms of the quality of the melodic filling, among other features. In order to help the reader understand this point, in Figures 1 and 2 we show a transcription of two version of the same *cante* to Western musical notation. A flamenco aficionado recognizes both versions as the same *cante* because certain notes appear in a certain order (they are called main notes). What happens between two of those notes does not matter regarding style classification, but does matter for assessing a performance or the piece itself. The main notes that the aficionado must hear have been highlighted in both Figures.

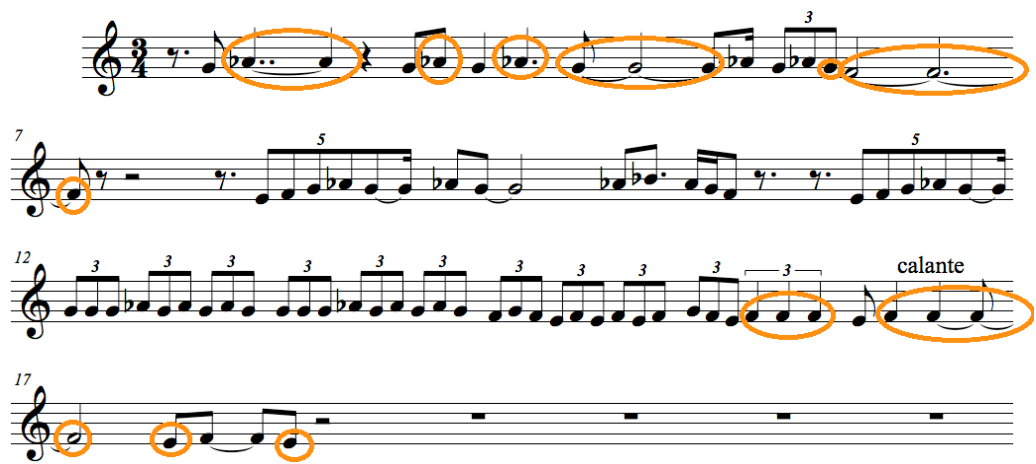

Figure 1: A *debla* by Antonio de Mairena.

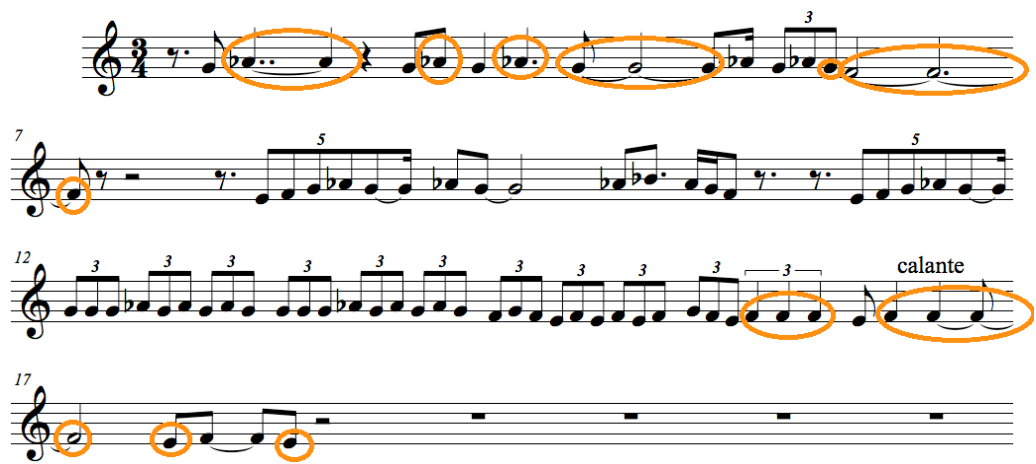

Figure 2: A *debla* by Chano Lobato.

In these transcriptions many melismas from the recording were removed for ease of reading. What is displayed constitutes an approximation to the performances; our point here is to illustrate how disparate two versions of the same *cante* may be. Furthermore, notice several of the features mentioned above (conjunct degrees, short *tessitura*, type of scale).

**2.4 The style of *tonás***

According to Blas Vega and Ríos Ruiz (1988) *tonás* (derived etymologically from the Spanish word *tonada,* air) arose from primitive Spanish folk songs adapted by flamenco singers in the early nineteenth century. Traditionally, the group of *tonás* includes *cantes* such as *martinetes, deblas, saetas, tonás*, and *carceleras*. These *cantes* were originally *tonás*, but later on they received particular names depending on other circumstances. For example, a *martinete* (a word etymologically close to hammer) is a kind of *toná* developed at smithies, and a *carcelera* is a *toná* whose subject matter is about prison. Since in flamenco music the word *toná* refers to both the style and one of its substyles, we will refer to "*tonás* style" as the whole style and *toná* as the substyle.

*Tonás* are *cantes* sung in free rhythm (occasionally, the pattern of *seguiriya* is used as rhythmic accompaniment). Each singer chooses his or her own reference pitch. Scale and melody type are modal. Frequent modes are major, minor, or Phrygian, though alternation of modes is also common (Fernández, 2004). The lyrics of these songs range widely. A classification of the *tonás* style mainly based on lyrics was carried out by Lefranc (2000). Blas Vega (1967) also studied the *tonás* style from a historical standpoint.

# 3 Melodic representation

## 3.1 Flamenco and its musical transcription

So far, flamenco music has been transmitted and preserved through oral tradition. Until very recently transcriptions have been scant and scattered. Because the guitar is a fixed-pitch instrument, Western notation has been employed to transcribe flamenco guitar music; see Hoces' thesis (2011) and the references therein for a thorough study on guitar transcription.

However, in the case of flamenco singing the situation is comparatively worse. There have been some attempts to use Western notation to represent flamenco, which have been proved fruitless for styles such as a cappella *cantes*. Only music with a strong metric structure and strict tuning seems to fit that kind of notation. Furthermore, a serious problem is the notation of flamenco singing techniques such as breathiness or nasalization in the voice. In spite of this situation, some transcription models have been proposed. Donnier (1997) proposed the adaptation of plainchant neumes to transcription of flamenco music. Hurtado and Hurtado (1998, 2002), on the contrary, forcefully argue for the use of Western notation. Disagreement exists over the most adequate and proper transcription methodology.

The problem of transcription in flamenco music would require further investigation, which is outside the scope of this paper. In this study, we used an automatic transcription method (Section 3.2) to extract a melodic transcription from a recording. After discussing with flamenco experts, we adopted the following transcription format. First, we used an equal-tempered scale for transcription, so that note pitches were quantized to an equal-tempered chromatic scale with respect to an estimated tuning frequency. Second, since we were analyzing musical phrases, we assumed a constant tuning frequency value for each excerpt. Third, even if the singer was out of tune we tried to approximate the used scale to a chromatic scale (mistuning was not transcribed). Next, we transcribed all perceptible notes, including short ornamentations in order to cover both expressive nuances and the overall melodic contour. Finally, the obtained transcription was post-processed to obtain a refined melodic contour holding the relevant information. In terms of format, the output of this process is a MIDI-like symbolic representation of the *cante*. The procedure for automatic transcription is presented below.

## 3.2 Automatic melodic transcription

### 3.2.1 Background

Given the lack of symbolic transcriptions of flamenco music, we had no choice but to only work with audio recordings. From recordings, automatic transcription systems compute a symbolic musical representation (Klapuri, 2006). For monophonic music material, the transcription thus obtained mainly preserves melodic features; in polyphonic music the central problem is to transcribe the predominant melodic line. Although existing systems provide satisfying results for a great variety of musical instruments, singing voice is still one of the most difficult instruments to transcribe, even in a monophonic context (Klapuri, 2006).

Current systems for melodic transcription are usually structured into three different stages, as represented in Figure 3: low-level, frame-based descriptor extraction (e.g., energy and fundamental frequency), note segmentation (based on location of note onsets), and note labelling.

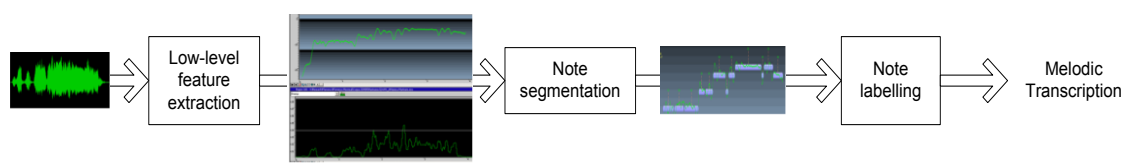

Figure 3: Stages in automatic melodic transcription.

### 3.2.2. Selected approach

The approach used in this study is summarized in Figure 4.

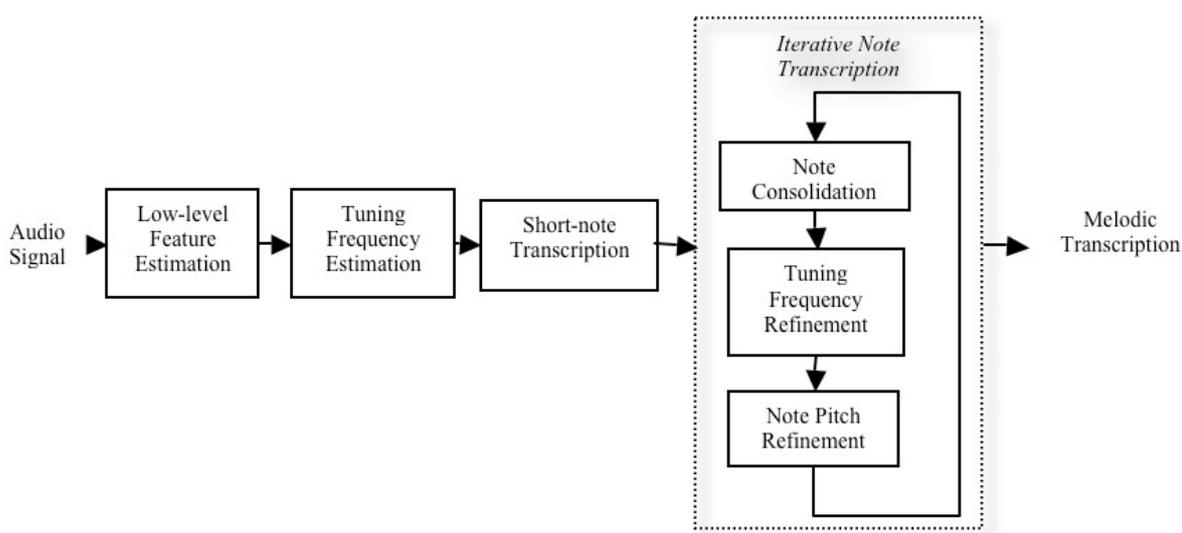

Figure 4: Diagram for melodic transcription.

The audio signal is first cut into frames of 23.2 ms each by using a frame rate of 60 frames per second. From each analysis frame, the spectrum is computed by using a 10-milisecond window. By following a frame-by-frame procedure, energy is computed and fundamental frequency (f0) estimated. The fundamental frequency estimation algorithm is based on the computation of amplitude correlation in the frequency domain.

From f0 and energy, an iterative approach for note segmentation and labelling is used, which consists of the following steps:

- Tuning frequency estimation. Since we are analyzing singing voice performances, the reference frequency (with respect to 440 Hz) is unknown. In order to locate the main pitches, an initial estimation of the tuning frequency (i.e. the reference frequency used by the singer to tune the piece) was made; an equal-tempered scale system was assumed. This tuning frequency is computed by minimizing the estimated instantaneous pitch error weighted average. The weights are computed by combining energy and first and second pitch derivatives.
- Short note transcription: the audio signal was then segmented into short notes by using a dynamic programming algorithm based on finding the segmentation that maximizes a set of probability functions. Those functions considered pitch error, energy variations and note durations.
- Iterative note consolidation and tuning frequency refinement: the estimated tuning frequency was then refined according to the obtained notes. In order to do this, the note pitch error weighted average was minimized, letting weights depend on note durations. Then, existing consecutive notes with the same pitch and a soft transition between them are consolidated. This process was repeated until there was no further consolidation.

This method has been tested against manual transcriptions on a flamenco music collection, including a variety of singers and recording conditions. We obtained an overall accuracy of 82% (100 cents tolerance) and 70% (50 cents tolerance) for a cappella singing, and we observed that transcription errors appear for noisy recordings, rough and detuned voices in highly ornamented sections. The reader is referred to Gómez and Bonada (2008, 2013) for further details.

The output of this step is then a symbolic representation of note pitch and duration. As a post-processing step, very short notes were detected and consolidated with their closest long note, and pitch values were converted into interval values for similarity computation.

## 4 Melodic similarity

As Pampalk et al. (2005) have pointed out, "unfortunately, music similarity is very complex, multi-dimensional, context-dependent, and ill-defined." Music similarity is complex because of music's inherent, quintessential complexity. To reflect that complexity, an ideal similarity measure should necessarily include low-, mid-, and high-level descriptors, as discussed in the introduction. Due to the scope of this paper, we do not take into account high-level descriptors (mood, motor and affective responses, etc.), which will be reserved for future work. Using only low-level descriptors in the design of music similarity measures has proved limited. Aucouturier et Pachet (2002) ascertain the existence of a glass ceiling that cannot be penetrated without integrating higher level cognitive variables into the measure; other authors, such as Pampalk et al. (2005), when studying similarity through timbre, have confirmed the presence of this glass ceiling. Recall that one of the main goals of this paper is to study melodic similarity in flamenco a cappella *cantes*. At a smaller scale, we have encountered the same difficulties as in the study of music similarity as a whole.

In an early work (Cabrera et al., 2008), we performed an analysis of melodic similarity of flamenco a cappella *cantes* by only examining melody, as represented by a sequence of note pitch and duration values. This generic melodic contour representation has been extensively used in the literature. For example, Suyoto and Uitdenbogerd (2008) studied the effect of using pitch and duration for symbolic music retrieval; van Kranenburg, Volk et al. (2009) looked into the problem of incorporating musical knowledge within alignment algorithms; Urbano et al. (2011) presented an geometric approach to musical similarity based on splines. Our results, although interesting and to a certain extent promising, revealed serious limitations, notably when a large number of *cantes* were analyzed or their variability was high. We understood the compelling need to incorporate specific descriptors into the design of our melodic similarity measure. Such descriptors had not yet been identified for these a cappella *cantes* to the best of our knowledge.

We then collaborated with flamenco experts in the identification and description of a set of mid-level descriptors, specific to the musical repertoire under consideration. This responds to the criticism expressed at the outset with respect to the collaboration between ethnomusicologists and technicians. Based on that set of descriptors we designed a melodic similarity measure. Again, only using mid-level specific descriptors would be as questionable as only using melodic contour (note pitch and duration). In fact, just for the sake of logical completeness, we carried out a separate analysis of a melodic similarity only grounded on mid-level descriptors. As expected, we found limitations, although of a different kind from those found when only melody was used. The limitations of both approaches are discussed later in this paper.

In parallel, we studied the mechanisms involved in the perception of melodic similarity in flamenco a cappella singing by contrasting human judgments of similarity for performances of a particular flamenco style, *martinete* (Kroher et al., 2014). We observed significant differences between the criteria used by non-expert musicians, who relied on surface features such as intervals, contour or timing, and flamenco experts, who performed a more in-depth structural analysis in terms of segmentation and symmetry. Nevertheless, we found significant correlation between their judgments both for synthetic and real melodies.

Following our findings, we propose to integrate both similarity measures (generic or melodic contour and specific mid-level features) into one global measure. The integrated measure allows greater accuracy, robustness, and musical sense.

We will then consider two distances: the first is the melodic contour distance (from now on MC distance), which measures variations on pitch and time duration; the second is the mid-level musical descriptors distance (from now on MD distance), which measures the distance between *cantes* based on mid-level descriptors.

# 5 Integrated approach to melodic similarity

## 5.1 Music collection

To start with, we gathered a set of 365 pieces belonging to the *tonás* style. Somehow, these *cantes* have scarcely been recorded compared to other *cantes*. Their solemn mood and emotiveness might be a plausible reason for that shortage of recordings. In spite of that, we spared no effort to gather as many recordings as possible from all feasible sources (private collections, libraries, historical recordings, several institutions, etc.). We may safely state that our collection is quite representative of this type of *cantes*. After the analysis phase of building the corpus, we decided to focus on three substyles, *deblas*, *martinete 1* and *martinete 2*. We came to that decision because of the following considerations: (1) The three styles are central to flamenco music; (2) We had information about singers, geographical locations, singing schools, dates, etc., which allows us to have a complete, in-depth characterization of them from a more general standpoint than just the musical one; (3) In general, our recordings have acceptable quality to carry out this and future research, which includes, for instance, automatic feature extraction from audio files; (4) There was a high number of recordings, 72 *cantes* in total, where 16 were *deblas* and 56 *martinetes* of type *1* and *2*; (5) Apart from the number, there was enough variability in the sample to test our methods; (6) These are styles have no accompaniment and this facilitates its automatic analysis. *Cantes* were labelled according to the metadata contained in the recording and also based on expert's criteria (sometimes the recording labels were incorrect). The 72 *cantes* chosen for the study were all *deblas, martinetes 1* and *2* included in the whole corpus. The corpus used for this study can be downloaded at http://mtg.upf.edu/download/datasets/tonas.

The musical features of *deblas*, *martinete 1*, and *martinete 2*, to be described below, were obtained after a thorough study. We carried out a set of interviews with a group of flamenco experts from Seville. First, we opened an analysis phase to identify which musical features were relevant to the characterization of the chosen *cantes*. Preliminary analysis produced too many variables or just variables with little explanatory power. Second, in search for the least complex yet meaningful description of *cantes*, we removed several variables. Most of the features identified were related to melody and form. The musical features were established out of the first phrase in the exposition, which was manually annotated by the flamenco experts.

## 5.2 Melodic dissimilarity distance

Although there is a great abundance of similarity measures proposed in the literature, many of them suffer from a common deficiency: lack of perceptual validity, i.e., they have not been tested on subjects. As a matter of fact, experiments with subjects are expensive and complex to carry out. Müllensiefen and Frieler (2004) tackled this problem head-on. First, they tried to establish some ground truth for melodic similarity under certain conditions; secondly, they analyzed 34 similarity measures (or dissimilarity distances) found in the existing literature to determine the most adequate measures in terms of perceptual validity.

These two authors conducted several experiments to build the desired ground truth. Previous efforts to build such ground truth were devoted by a number of authors such as Schmuckler (1999), McAdams and Matzkin (2001), and Pardo et al. (2004), but their attempts proved insufficient. Müllensiefen and Frieler paid a great deal of attention to selecting those similarity measures that best approximate the similarity of human music experts (their experiments were conducted on music experts, given that subjects with little or no music background showed great inconsistency).

Given two similarity measures, any linear combination of them will result in another similarity measure. This possibility was also taken into consideration by Müllensiefen and Frieler, who, once the ground truth was obtained, modeled subject's ratings with linear regression. They concluded that the best similarity measure $\sigma_{best}$ is

$$\sigma_{best} = 3.355 \cdot rawedw + 2.852 \cdot ngrcoord,$$

where *rawedw* is the rhythmically weighted raw pitch edit distance, and *ngrcoord* is the count distinct measure. See their paper for more details.

In our work we followed the methodology proposed by Müllensiefen and Frieler. We were aware that their ground truth was established for Western music. For want of a better ground truth, we decided to use this one, as we had no other choice. We measured the melodic similarity of all *cantes* in our collection by using the measure above. The measure was applied to the transcriptions output by the algorithm described in the previous Section; recall that the output of the transcription algorithm is in symbolic format. Once the weighted raw pitch edit distance and the count distinct measure are obtained, the final similarity value is the linear combination of both with the weights given in the formula above. For the actual implementation of the melodic similarity algorithms we used the open source library *SimMetrics* (Chapman, 2006).

### 5.3 Mid-level dissimilarity distance

In this Section the mid-level dissimilarity distance, or simply the mid-level distance, is introduced. As described above, the mid-level distance is based on musicological features of the given *cantes*. As a first step, we characterized the main musical features of the three styles under consideration. This characterization, which is a piece of pure musicological research, is valuable on its own. To our knowledge no description of these a cappella *cantes* in these terms has been provided before. After describing the features of the *cantes*, we proceeded to extract a set of common features to the three styles. Based on this set of features we designed the mid-level distance.

#### 5.3.1 Musical features of *deblas*

The *debla* is a song from the style of *tonás*. In general, it is marked by its great melismatic ornamentation, more abrupt than the other songs from this style, which characterizes its melody. The musical features that characterize the different variants within the *debla* style are the following.

1. **Beginning by the word *¡Ay!***: *¡Ay!* is an interjection expressing pain. This is quite idiosyncratic to flamenco music, as its presence in a *cante* is a distinguishing feature. Values of the variable: Yes and no.

2. **Linking of *¡Ay!* to the text.** That initial *¡Ay!* may be linked to the text or just be separated from it. Values of the variable: Yes and no.
3. **Initial note.** It refers to the first note of the strophe. Normally, it is the sixth degree of the scale, but sometimes the fifth degree also appears. Values of the variable: 5 and 6 (no other value appears).
4. **Direction of melody movement in the first hemistich.** (A hemistich is half of a line of a verse.) The direction can be descending, symmetric, or ascending. The detection of the melody movement has to be performed irrespective of the melismas found between the main notes; see the discussion held in Section 2.3.
    - Descending movement: When the movement is descending, the melody starts off with the sixth degree and it develops by gradually descending to the fourth degree. It is considered as descending in direction when there is a quick initial appoggiatura from the fifth degree to the sixth followed by a fall to the fourth degree.
    - Symmetric movement: The first hemistich begins with a rise from the third degree to the sixth, and then it falls to the fourth degree.
    - Ascending: When the direction of the melody is just ascending.

    Values here are *D*, *S,* and *A*.

5. **Repetition of the first hemistich.** That repetition may be of the whole hemistich or just a part of it. Values of the variable: Yes or no.
6. **Caesura.** The caesura is a pause that breaks up a verse into two hemistiches. Values of the variable: Yes and no.
7. **Direction of melody movement in the second hemistich.** It has the same description as in the first hemistich. Values of the variable are *D*, *S,* and *A*.
8. **Highest degree in the second hemistich.** It is the highest degree of the scale reached in the second hemistich. Usually, the seventh degree is reached, but fifth and sixth degrees may also appear. Values of the variable are 5, 6, and 7.
9. **Frequency of the highest degree in the second hemistich.** The commonest melodic line to reach the highest degree of the scale consists of the concatenation of two *torculus* (a three-note neume where the central note is higher in pitch than the other two notes). The value of this variable indicates how many times this neume is repeated in the second hemistich.
10. **Duration.** Although the duration is measured in milliseconds, our intention was to classify the *cantes* into three categories, fast, regular, and slow. To do so, we first computed the average $\mu$ and the standard deviation $\sigma$ of the durations of all the *cantes* in the music collection. Then, fast *cantes* are those whose duration is less than $\mu-\sigma$, regular *cantes* have their duration in the interval $[\mu-\sigma, \mu+\sigma]$, and slow *cantes* have durations greater than $\mu+\sigma$. Values of this variable are *F, R,* and *S*.

### 5.3.2 Musical features of *martinetes*

There are three main styles, which are named *martinetes* in the flamenco literature. The first one, to be called *martinete* 1, has no introduction, whereas the second one, to be called *martinete* 2, mostly starts with a couple of verses from a *toná*. The third one, to be called *martinete* 3, is a concatenation of a *toná* and some of the previous variants of *martinetes*; the *toná* of *martinetes* 2 and 3 is called *toná* incipit. Because *martinete 3* is a combination of *toná* and *martinetes 1* and *2,* we removed it from our current study, as we just sought characterizing the most fundamental styles.

The musical features of the *martinete 1* are the following.

1. **Repetition of the first hemistich.** As in the case of *deblas* repetition may be complete or partial. Values of the variable: Yes and no.
2. **Clivis (or flexa) at the end of the first hemistich.** Normally, fall IV-III or IV-IIIb are found (again this is detected irrespective of melismas). The commonest ending for a strophe is the fourth degree, whose sound is sustained until reaching the caesura. Some singers like to end on III or IIIb. Values of the variable are: Yes and no.
3. **Highest degree in both hemistichs.** The customary practice is to reach the fourth degree; some singers reach the fifth degree. Values of the variable are 4 and 5.
4. **Frequency of the highest degree in the second hemistich.** The melodic line is formed by a *torculus,* a three-note neume, III-IV-III in this case. This variable stores the number of repetitions of this neume.
5. **Final note of the second hemistich.** The second hemistich of *martinete 1* is ended by falling on the second degree. Sometimes, the second degree is flattened, which produces Phrygian echoes in the cadence. This variable takes two values, *1* when the final note is the tonic and *2* when the final note is II.
6. **Duration.** This variable is defined as in the case of *deblas* (that is, in terms of $\mu$ and $\sigma$). Values of this variable are *F, R,* and *S*.

As for *martinete 2* we have the following features.

1. **Highest degree in both hemistichs.** In this case the customary practice is to reach the sixth degree; in some cases singers just reach the fourth or fifth degrees. Values of the variable are 4, 5, and 6.
2. **Frequency of the highest degree in the second hemistich.** In this case the neume is also a *torculus*. This variable stores the number of repetitions of this neume.
3. **Symmetry of the highest degree in the second hemistich.** The second hemistich of a *martinete 2* is rich in melismas. This feature describes the distribution of the melismas around the highest reached degree, which is usually the sixth degree. Melismas can occur before and after reaching the highest degree (symmetric distribution), only before the highest degree (left asymmetry) or only after the highest degree (right asymmetry). Values of the variable are *S, L,* and *R*.
4. **Duration.** This variable is defined as in the previous cases. Values of this variable are *F, R,* and *S*.

### 5.3.3 Common features

Carrying out the preceding analysis allowed us to extract a set of musical features to be used in the definition of musical similarity between *cantes*. As a matter of fact, just using features very peculiar to a given style would distort the analysis, as their discriminating power would be very high. Our intention was to select a set of a few features capable of discriminating between different *cantes*. The final set of variables was the following.

1. Initial note of the piece;
2. Highest degree in both hemistichs;
3. Symmetry of the highest degree in the second hemistich;
4. Frequency of the highest degree in the second hemistich;
5. Clivis at the end of the second hemistich;
6. Final note on the second hemistich;
7. Duration of the *cante*.

Note that some of these variables do not appear as features in some *cantes*; for example, clivis is not a feature of *deblas*. In order to avoid style-specific variables, which would distort the power of the distance, we removed those variables that only accounted for only one *cante*. In the case of clivis, this featured remained, as it was present in the description of *martinete 1* and *martinete 2*.

The distance we used to measure the dissimilarity between two *cantes* was the simplest one could think of, the Euclidean distance. We just computed the Euclidean distance between features vectors. Our intention was to test how powerful the musical features would be. The Euclidean is just a geometrical distance and does not reflect perceptual distance whatsoever. However, because of the robustness and power of the musical features, results were good.

**5.4 Integrated distance**

Once the MC and MD dissimilarity distances were obtained, the next task was to propose a reasonable manner to integrate both into one distance $d_I$. Guided by the principle of simplicity, we proposed a linear combination of the MC and MD distances as integrated distance,

$$d_I = (1-\alpha) \cdot d_{MC} + \alpha \cdot d_{MD},$$

where $\alpha$ is a coefficient to be determined.

In order to integrate different music similarity measures, Schedl et al. (2011) mentioned the possibility of letting users control the weight $\alpha$ for different distance measures or criteria. This would ideal to define user-adapted metrics (for example, for naïve listener vs flamenco experts), but it would require great time and effort from the user to make her preference explicit. For this reason, this approach is usually adopted for small datasets, as in Kroher et al. (2014), where user ratings are gathered for 11 versions of the same flamenco style.

An alternative approach to evaluating similarity measures is to relate similarity to categorization (Berenzweig et al. 2003), which allows dealing with larger music collections. This is the approach followed in this study, where we associate similarity to categorization, and then refine the weight of the different measures according to how well they can separate the different styles.

**5.5 Distance assessment**

As mentioned below, we assessed the melodic contour distance by running a classification experiment. In this classification experiment, we classified *cantes* by employing the nearest centroid classifier (Manning et al. 2008). Again, we insist that the ultimate goal of this paper is not to design a distance to carry out classification tasks per se, but to determine the value of $\alpha$ and explore the behavior of the distance. Our distance is not designed to be part of recommender systems, or music automatic categorization systems, or the like. Suitable precision, recall, and f-score measures were computed for that classifier. We complemented these measures with a clustering analysis carried out through phylogenetic techniques. Furthermore, we addressed the issue of how to choose coefficient $\alpha$ for the integrated distance by performing that classification task.

### 5.5.1 Performance measures for MC distance in style classification

*Cantes* are classified as follows. A *cante* is classified according to the style of its nearest centroid (mean) as measured by the MC distance. Thus, first compute the centroids of each style (*debla, martinete 1*, and *martinete 2*), where the *cante* to be classified is excluded. Then classify the *cante* as belonging to the style of the nearest centroid. Every *cante* is classified in this manner and this procedure produces classification results each entire style.

Table 1 below summarizes the relevant values of the measures, which are briefly reviewed next. Let $t_p$ be the number of true positives (*cantes* correctly classified), and $f_p$ the number of false positives (*cantes* incorrectly classified). Precision $P$ is defined as the ratio $\frac{t_p}{t_p + f_p}$. False negatives, $f_n$, are missing results (*cantes* not appearing in the classification of a style). Recall $R$ is defined as the ratio $\frac{t_p}{t_p + f_n}$. The f-score measure is the harmonic mean of precision and recall, $2\frac{P \cdot R}{P + R}$. In the microaverage method a global measure is computed out of values $t_p$, $f_p$, and $f_n$ for all styles; for example, microaverage precision is the sum of all true positives over all styles divided by the sum of all true positives plus false positives also over all styles. Macroaverage is simply the mean of the values obtained for each style. For further details, see Manning et al. (2008).

|  | *Martinete 1* | *Martinete 2* | *Debla* | Micro-average | Macro-average |
| --- | --- | --- | --- | --- | --- |
| **Precision** | 0.77 | 0.8 | 0.51 | 0.68 | 0.69 |
| **Recall** | 0.75 | 0.4 | 0.87 | 0.68 | 0.67 |
| **F-score** | 0.76 | 0.53 | 0.65 | 0.89 | 0.68 |

Table 1: Main performance measures for the melodic contour distance.

In order to ensure that the mean was representative of the *cantes*, the coefficient of variation was computed. Its value was below 15%, which indicated low dispersion around the mean. Classification results for *martinete 1* present little variation. In the case of *martinete 2* precision is twice the value of recall. This implies there are a high number of *cantes* not appearing in the classification of this style. Contrary to *debla*, precision is relatively low (many misclassified *cantes*) and recall is high. Microaverage precision and recall have the same value, 0.68, which gives an overall idea of the classification results. The conclusion is that the melodic contour distance does not perform well on these styles of flamenco *cantes*.

**5.5.2 Cross validation for mid-level variables**

Some of the mid-level variables are very specific as can be seen from their description; furthermore, there are a high number of mid-level variables. These two facts could give the impression that there might be an overfitting issue and cast doubt on the validity of the MD distance. In order to dispel any doubts, a 10-fold cross validation test was carried out. Partition of the dataset was done randomly. The variables used in the cross validation test were the following: (1) Initial note; (2) Symmetry of the highest degree in the second hemistich; (3) Frequency of the highest degree in the second hemistich; (4) Clivis at the end of the second hemistich; (5) Final note on the second hemistich; (6) Highest degree in the *cante*; (7) Duration of the *cante*. The 10-fold cross validation was carried through linear discriminant analysis. In our case, this analysis provides two discriminant functions, which are linear combinations of the previous variables. Those two functions are then used as classifiers for each step of the cross validation. The functions are computed so that groups (*cantes*) are maximally separate in the most parsimonious way (that is, by minimizing the number of variables involved).

Our data met the assumptions necessary for performing the linear discriminant analysis. The *p*-value for the test of equality of means was less than 0.05 in all cases. Tests for homogeneity of covariance matrices and equal variance proved positive. Moreover, all independent variables had discriminatory power (as confirmed by Wilk's lambda). Wilk's lambda values associated with the test of discriminant functions were less than 0,001, which points to a good discriminatory ability of the two functions.

All *cantes* but one were correctly classified in the cross validation (a *martinete 1* was classified as a *martinete 2*). This proves that the mid-level descriptors identified here are suitable for describing flamenco *tonás*.

At this point, it seems there is an apparent contradiction, for if the classification performed by using mid-level variables is so accurate, why then considering low-level variables? That contradiction is resolved when we realize that the computation of some mid-level variables actually involves measuring values in the presence of complex ornamentations. For example, the clivis (or flexa) is a mid-level variable utilized in the description of *martinete 1*. It takes value 'yes' if there is a fall at the end of the first hemistich. Between the two notes forming that fall there can be all type of ornamentations, often very complex and rich, and still the variable would take value 'yes'. We discussed this phenomenon in depth in Section 2.3. Therefore, whereas the low-level variables were computed by purely computational methods, mid-level variables needed manual annotation –very human annotation would be a fairer description. As a matter of fact, the annotation of these variables by human beings included complex cognitive skills in terms of musical pattern recognition. In order to bring about an unbiased situation, we repeated the computation of mid-level variables removing those requiring purely human annotation. The final set of variables was composed of the following four variables: initial note of the piece; highest degree in both hemistichs; final note on the second hemistich; duration of the *cante*. With this new set of four variables, we performed a 10-fold cross validation again. Out of 72 *cantes*, there were 8 misclassified *cantes*, an 11.11% of the total; Cronbach's alpha was very close to 1.0 (above 0.9). Results were reasonably good given the final set of variables used. We will carry out further analysis of the mid-level in the next Section, once phylogenetic graphs are introduced.

### 5.5.3 Clustering and phylogenetic graphs

Distance matrices can be better visualized by employing phylogenetic graphs (Huson and Bryant, 2006), a visualization technique borrowed from Bioinformatics. Given a distance matrix from a set of objects, a phylogenetic graph is a graph whose nodes are the objects in the set and such that the distance between two nodes in the graph corresponds to the distance in the matrix. Obviously, this property cannot be held for arbitrary matrices. The phylogenetic graph algorithm provides an index, the *LSFit*, expressed as a percentage. This index indicates how accurate the correspondence between the distances in the graph and the distances in the set of objects is. The higher the index is, the more accurate the correspondence between matrix and graph distances is. To actually compute our phylogenetic graphs we used SplitsTree, an implementation by Huson and Bryant (2006). In general, clustering and other properties are easier to visualize with phylogenetic trees.

In Figure 5 the phylogenetic graph corresponding to $d_{MC}$ is depicted. Three clusters can be discriminated, which roughly matches the three styles. Although the clustering is in general correct, the graph suffers from a poor resolution. The *LSFit* for this graph is 99.19%.

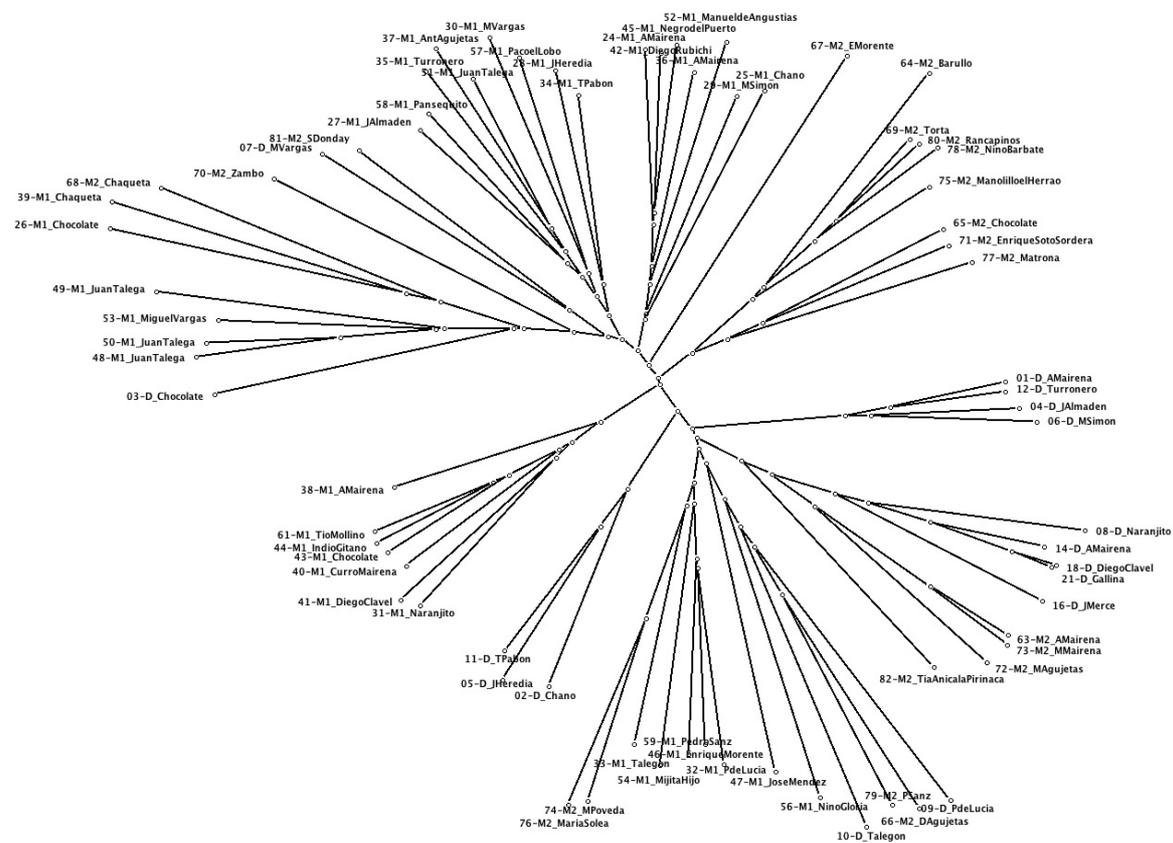

Figure 5: The phylogenetic graph for the MC distance.

Figure 7 displays the phylogenetic graph for distance $d_{MD}$ with the seven variables (that is, those described in Section 5.3.3). Label D stands for *debla*, label M1 stands for *martinete 1* and label M2 for *martinete 2*. The *LSFit* for this graph is 98.36%. The graph highlights more complex relations among the *cantes*. In this graph we can appreciate three clusters, one for *martinete 1*, located in the bottom of Figure 6; one for *martinete 2*, around the upper left corner; and one for *debla* located around the upper right corner. Within each cluster there are smaller clusters that show differences between performances. Since for this graph we used the full set of variables, the discriminatory power is very high, as predicted by the above computations in Section 5.5.2 (cross validation).

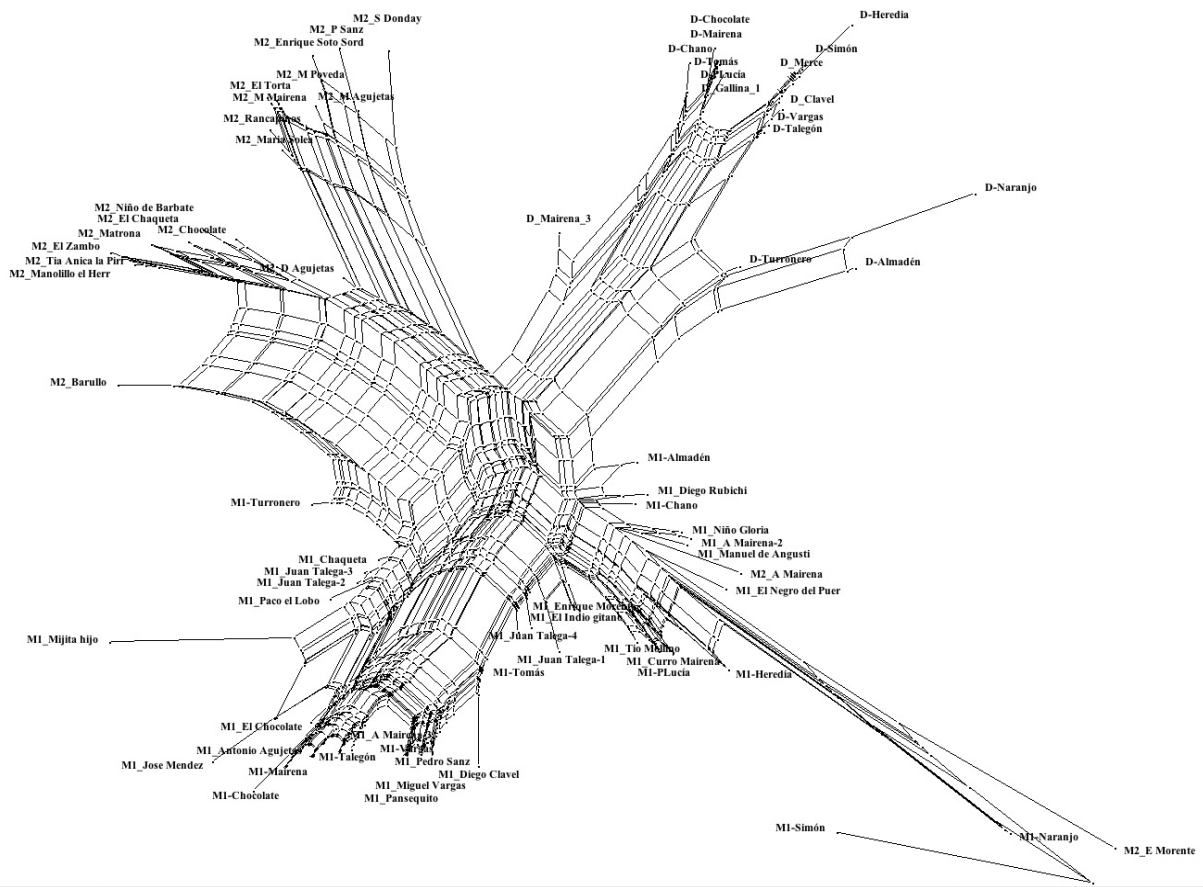

Figure 6: The phylogenetic graph for the MD distance with the full set of variables.

Figure 7 shows the phylogenetic graph for the reduced set of mid-level variables; the *LSFit* for this graph is 94.12%. There is a fact that immediately captures the reader's attention: There are *zero distances* among some *cantes*, which stand out as nodes with more than one *cante* associated with it. To certain extent, this fact should not surprise. The most style-specific mid-level variables were removed, which also turned out to be those that could not be automatically computed. However strange this situation may sound, it provides a solid rationale to warrant a combined distance between low-level and mid-level distances, especially when both distances are intended to be computed automatically.

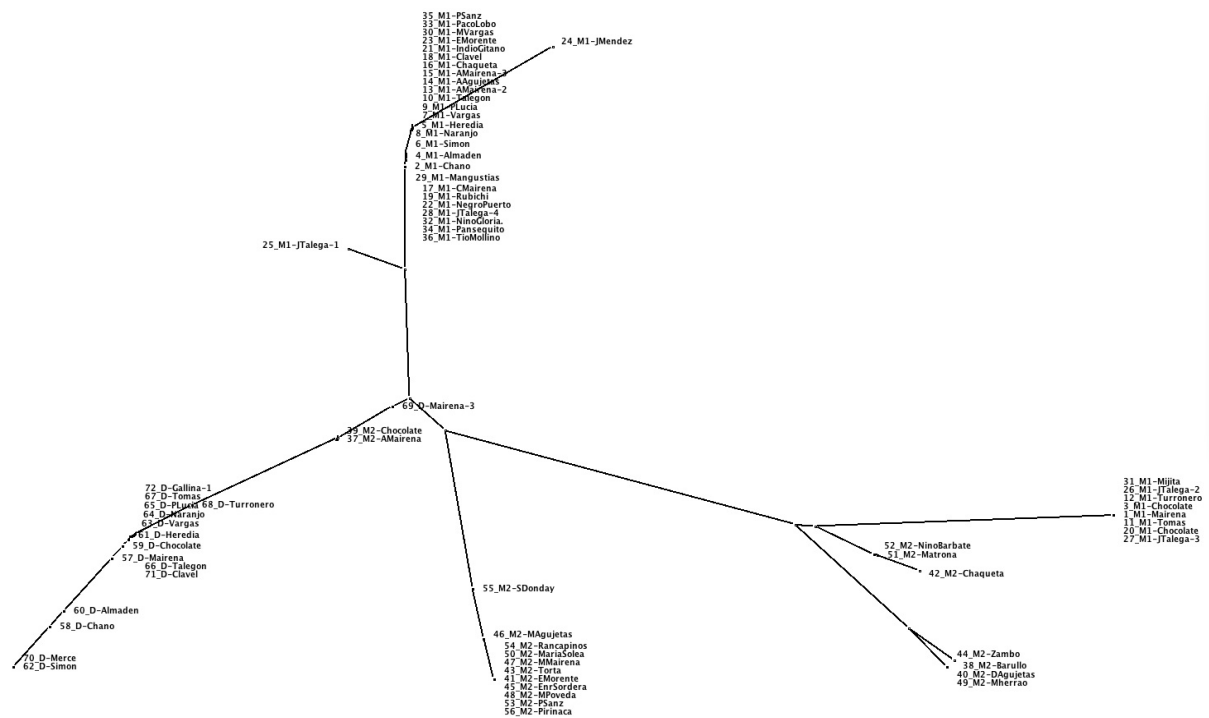

Figure 7: The phylogenetic graph for the MD distance with the reduced set of variables.

### 5.5.4 Analysis of coefficient $\alpha$ in the integrated distance

It remains the problem of selecting a value for $\alpha$ in the formula of the integrated distance. The ideal manner to do it would be through experiments with subjects. However, at the present stage of this work, experiments were not possible to conduct. Therefore, in order to determine the value of $\alpha$, we carried out a classification task. We classified the set of *cantes* by using a *k*-nearest neighbours classifier and by taking the reduced set of variables for the MD distance. Given a *cante,* if its *k* nearest neighbours belong to the same style, the *cante* will be classified as belong to that style. A *cante* is missclassified if the labelling is incorrect or all the neighbours do not belong to one style. With regard to the choice of parameter *k*, we followed the work of Duda et al. (2012), where they contend that a good value for *k* is the floor of the square root of the number of the objects. In our case, we have 72 *cantes,* 36 being *martinete 1,* 20 being *martinete 2*, and 16 being *deblas*. The number of neighbours for each case is then, respectively, 6, 4, and 4. We recall that the combined distance is defined as $d_I = (1-\alpha) \cdot d_{MC} + \alpha \cdot d_{MD}$, where $\alpha$ ranges from 0 to 1. Because of the difference in range, both distances were normalized before combining them into the $d_I$. In Table 2 below it is shown both the number of misclassified *cantes* as well as the performance measures for the three styles.

| α | 0 | 0.1 | 0.2 | 0.3 | 0.4 | 0.5 |
|---|---|---|---|---|---|---|
| **Number of misclassified *cantes*** | 40 | 10 | 7 | 11 | 12 | 12 |
| **Percentage of misclassified cantes (%)** | 55,55 | 13,88 | 9,72 | 15,27 | 16,66 | 16,66 |
| **Precision – M1** | **89,29** | 100,00 | 100,00 | 100,00 | 100,00 | **97,14** |
| **Recall – M1** | 30,56 | 2,78 | 2,78 | 2,78 | 2,78 | 5,56 |
| **f-score – M1** | 45,53 | 5,41 | 5,41 | 5,41 | 5,41 | 10,51 |
| **Precision – M2** | 100,00 | 100,00 | **93,75** | 100,00 | 100,00 | **91,67** |
| **Recall – M2** | 84,21 | 36,84 | 21,05 | 47,37 | 47,37 | 42,11 |
| **f-score – M2** | 91,43 | 53,85 | 34,38 | 64,29 | 64,29 | 57,70 |
| **Precision – D** | 100,00 | 100,00 | 100,00 | 100,00 | 100,00 | 100,00 |
| **Recall – D** | 76,47 | 11,76 | 11,76 | 5,88 | 11,76 | 17,65 |
| **f-score – D** | 86,67 | 21,05 | 21,05 | 11,11 | 21,05 | 30,00 |

| α | 0.6 | 0.7 | 0.8 | 0.9 | 1.0 |
|---|---|---|---|---|---|
| **Number of misclassified *cantes*** | 17 | 21 | 24 | 31 | 6 |
| **Percentage of misclassified cantes (%)** | 23,61 | 29,16 | 33,33 | 43,05 | 8,33 |
| **Precision – M1** | **94,29** | **96,55** | 93,55 | **86,67** | **97,22** |
| **Recall – M1** | 8,33 | 22,22 | 19,44 | 72,22 | 72,22 |
| **f-score – M1** | 15,31 | 36,13 | 32,20 | 78,79 | 82,88 |
| **Precision – M2** | **91,67** | 100,00 | **85,71** | **75,00** | 100,00 |
| **Recall – M2** | 57,89 | 57,89 | 68,42 | 68,42 | 84,21 |
| **f-score – M2** | 70,97 | 73,33 | 76,10 | 71,56 | 91,43 |
| **Precision – D** | 100,00 | 100,00 | 100,00 | 100,00 | 100,00 |
| **Recall – D** | 17,65 | 17,65 | 23,53 | 47,06 | 11,76 |
| **f-score – D** | 30,00 | 30,00 | 38,10 | 64,00 | 21,05 |

Table 2: Main performance measures for the integrated distance

The value for $\alpha$ that minimizes the total number of mistakes is 0.2, with 9.72% of mistakes. The number of mistakes first decreases as $\alpha$ increases until reaching $\alpha = 0.2$; after that point the number of mistakes steadily increases. Last column of Table 2 shows the information about distance $d_{MD}$. Note that as soon as $d_{MD}$ and $d_{MC}$ are combined, no two *cantes* have the same distance any more. The integrated distance possesses, therefore, a better discrimination power.

Furthermore, examining the rest of the performance measures provide an accurate picture of the behavior of distance $d_I$. Given the classification scheme used, a *cante* is either not classified (because its nearest neighbors do not belong to the same style) or is misclassified (because all its nearest neighbors belong to the incorrect style). Recall measures the former behavior whereas precision measures the latter. Cases where misclassified *cantes* are produced are shown in bold in Table 2. Most of these cases corresponded to *martinete 2* being classified as *martinete 1*, followed in frequency by the case of *debla* being classified as *martinete 2*. Worth remarking also is the fact that some particular *cantes* are consistently misclassified under different values of $\alpha$. As for the case of *cantes* not classified, we found that in most cases only one neighbor was different from the other three ones.

## 6 Conclusions and future work

The criticism expressed by Byrd and Crawford (2002) as well as Cornelis et al. (2010) is indeed telling and sharp. As has been repeated throughout this paper, we have tried to satisfy that criticism. A multidisciplinary team (including flamenco experts) carried out a study of flamenco music, in which the team proposed a melodic distance integrating general (automatically computed from melodic contour, $d_{MC}$) and specific (manually extracted by flamenco experts $d_{MD}$) descriptors. The differences between the behavior of $d_{MC}$ and $d_{MD}$ and $d_I$ can probably be explained better by turning to phylogenetic graphs. The graph associated with distance $d_{MC}$ showed a somewhat poor resolution, whereas the one associated with $d_{MD}$ had the zero-distance problem. The combination of distances $d_{MC}$ and $d_{MD}$ into $d_I$ solved both problems, lending validity to this approach to melodic similarity. The phylogenetic graph associated with $d_I$ displayed a good resolution with all positive distances.

The approach developed in this work, multidisciplinary study of flamenco melodic similarity, seems to bring good results. As already stated, melodic similarity is, "very complex, multi-dimensional, context-dependent, and ill-defined" (Pampalk et al., 2005). In our work we addressed this issue by presenting a melodic similarity distance coalescing $d_{MC}$ and $d_{MD}$ into $d_I$. Also, we would like to highlight the importance of the musical characterization of a cappella *cantes*, which led to the definition of $d_{MD}$. Such musical characterization –to the best of our knowledge- is original research.

The distance based on the mid-level descriptors was computed manually. This was due to a complex problem found in the automatic extraction of some mid-level descriptors. For example, let us look at the presence of clivis, the second musical feature in *martinete 1*. It just traces the presence of the fall IV-III, but without taking into account the melismas that may appear in that fall. An automatic system to compute $d_{MD}$ should detect melismas, discard them, and measure the corresponding feature. Unfortunately, for a cappella *cantes* such task is challenging and arduous. Melismas have no formal definition in these types of *cantes*. Since there is no meter, terms as strong and weak beats are not meaningless, and therefore they cannot be used in detecting melismas. Defining melismas in terms of note duration has as well proven fruitless (Gómez et al., 2011). In flamenco it is very common to repeat a melisma at different tempi, ranging from very fast to very slow, and therefore note duration is of no use. Check, for instance, the final melisma in Figure 1 to be fully aware of this problem. Other criteria, such as articulation, type of interval, final note have also proved unsuccessful. At the present time, the authors are actively researching this issue. Once this question is settled, an automatic system will be designed.

The work presented here can be extended in a number of ways. In the future we will study other a cappella *cantes* such as *saetas, carceleras,* or *nanas*. Also, we would like to consider *a cappella cantes* and polyphonic where the first musical phrase is not deciding in terms of style classification.

Another unresolved issue is the choice of $\alpha$, which ideally should be set through experiments with subjects. That will also be part of our future work.

## Acknowledgments

Tipo de Proyecto: Proyectos de Excelencia de la Junta de Andalucía
Referencia: P12-TIC-1362

## References

Aucouturier, J.-J., and Pachet, F. (2002). Music similarity measures: How high is the sky? *Proceedings of International Symposium on Music Information Retrieval*.

Berenzweig, A., Logan, B., Ellis, D., Whitman, B. A Large-scale Evaluation of Acoustic and Subjective Music Similarity Measures, Proceedings of International Conference for Music Information Retrieval, ISMIR, 2003.

Blas Vega, J. (1967). *Las tonás*, ed. Guadalhorce, Málaga.

Blas Vega, J. and Ríos Ruiz, M. (1988). *Diccionario enciclopédico ilustrado del flamenco*. Cinterco. Madrid.

Byrd, D. and Crawford, T. (2002). Problems of music information retrieval in the real world. *Information Processing and Management*. Pages 249-272.

Cabrera, J.J., Díaz-Bañez, J.M., Escobar-Borrego, F.J., Gómez, E., Mora, J. (2008). *Comparative Melodic Analysis of A Cappella Flamenco Cantes*. Fourth Conference on Interdisciplinary Musicology (CIM08), Thessaloniki, Greece, July 2008.

Castro Buendía, G. (2010) Los cantes sin guitarra en el flamenco, antecedentes musicales y modalidades. *Revista Universitaria de Investigación sobre Flamenco "La Madrugá"* Vol. 2.

Chapman, S. (2006) *SimMetrics*. Natural Language Processing Group. Sheffield University. http://staffwww.dcs.shef.ac.uk/people/sam.chapman@k-now.co.uk/simmetrics.html

Cornelis O., Lesaffre M., Moelants D., Leman M. (2010) Access to ethnic music: Advances and perspectives in content-based music information retrieval. *Signal Processing*. Pages 1008-1031.

Donnier, P. (1997). Flamenco: elementos para la transcripción del cante y la guitarra, *Proceedings of the III Congress of the Spanish Ethnomusicology Society*.

Duda, R.O., Hart, P.E., Stork, D.G. (2012) *Pattern classification*. John Wiley & Sons.

Fernández, L. (2004) *Teoría musical del flamenco*. Acordes Concert, S.L. Madrid. English version: *Flamenco Music Theory*.

Gamboa, J. M. (2005). *Una historia del flamenco*. Espasa-Calpe. Madrid.

Gómez, E., Bonada, J. (2008). *Automatic Melodic Transcription of Flamenco Singing*. Fourth Conference on Interdisciplinary Musicology (CIM08), Thessaloniki, Greece, July 2008.

Gómez, E and Bonada J. (2013). Towards Computer-Assisted Flamenco Transcription: An Experimental Comparison of Automatic Transcription Algorithms As Applied to A Cappella Singing. *Computer Music Journal*. 37(2), 73-90.

Gómez, P., Pikrakis, A., Mora, J., Díaz-Bañez, J.M., Gómez, E. (2011). *Automatic Detection of Ornamentation in Flamenco Music. Proceedings of the 4th International Workshop on Machine Learning and Music: Learning from Musical Structure*, Granada, Spain, December 2011.

Hoces, R. (2011) *La transcripción musical para guitarra flamenca: análisis e implementación metodológica*. Ph.D. thesis. Universidad de Sevilla.

Hurtado Torres, D. and Hurtado Torres A. (1998) *El arte de la escritura musical flamenca*. Bienal de Arte Flamenco. Sevilla.

Hurtado Torres, A. and Hurtado Torres, D. (2002). *La voz de la tierra, estudio y transcripción de los cantes campesinos en las provincias de Jaén y Córdoba*. Centro Andaluz de Flamenco. Jerez.

Huson, D.; and Bryant, D. (2006). Application of phylogenetic networks in evolutionary studies. *Molecular Biology and Evolution*, 23:254-67.

Klapuri, A. and Davy, M. (Editors). *Signal Processing Methods for Music Transcription*. Springer-Verlag, New York, 2006.

van Kranenburg, P. and Volk, A. and Wiering, F. and Veltkamp, R. C. (2009) *Musical Models for Folk-Song Melody Alignment*, *Proceedings of the International Society on Music Information Retrieval conference*, Kobe, Japan, 507-512.

Kroher, N., Gómez E., Guastavino, C., Bonada, J., Gómez, P. (2014) Computational models for perceived melodic similarity in a cappella flamenco cantes, *Proceedings of the 15th International Society for Music Information Retrieval Conference*, Taipei, Taiwan, October 2014.

Lefranc. P. (2000). *El Cante Jondo. Del territorio a los repertorios: tonás, seguiriyas, soleares*. Publicaciones de la Universidad de Sevilla, colección Cultura Viva, n° 16.

Manning, C.; Raghavan, P.; Schütze, H. (2008). Vector space classification. *Introduction to Information Retrieval*. Cambridge University Press.

McAdams, Stephen, and Matzkin, Daniel. (2001). Similarity, invariance, and musical variation in *The Biological Foundations of Music*, edited by Robert J. Zatorre and Isabelle Peretz (New York: New York Academy of Sciences), 62-74.

Molina, R. ; Mairena, A. (1963). *Mundo y formas del cante flamenco*. Librería Al-Andalus, 1963 (reprint by J. Cenizo, Sevilla: Giralda, 2004).

Müllensiefen, D., and Frieler, K. (2004). Cognitive adequacy in the measurement of melodic similarity: algorithmic vs. human judgments. *Computing in Musicology*, 13, 147-176.

Navarro, J.L. and Ropero, M. (editor) (1995). *Historia del flamenco*. Ed. Tartessos, Sevilla.

Pampalk, E.; Flexer, A.; and Widmer, G. (2005) Improvements of audio-based music similarity and genre classification. *Proceedings of the International Symposium on Music Information Retrieval.* Pages 628-633.

Pardo, B., Jonah S., and Birmingham, W. (2004). Name that tune: A pilot study in finding a melody from a sung query. *Journal of the American Society for Information Science and Technology* 55/4.

Schedl, M.; Stober, S.; Gómez, E.; Orio, N.; Liem, CCS (2011). User-Aware Music Retrieval. Multimodal Music Processing 3, 135-156

Schmuckler, M. A. (1999). Testing models of melodic contour similarity. *Music Perception.* Vol. 16, No. 3, 109-150.

Sundberg, J. (1987). The Science of the singing voice. DeKalb, IL: Northern Illinois Univ. Press.

Suyoto, I.S.H. and Uitdenbogerd, A.L. (2008). The effect of using pitch and duration for symbolic music retrieval. Proceedings of the 13th Australasian Document Computing Symposium, Hobart, Australia. December 8th, 2008.

Tzanetakis, G. A. Kapur, W. A. Schloss and M. Wright (2007). Computational ethnomusicology. *Journal of Interdisciplinary Music Studies,* vol. 1(2).

Urbano, J.; Lloréns, J.; Morato, J.; Sánchez-Cuadrado S. (2011). Melodic similarity through shape similarity. K. Jensen et al. (Eds.): CMMR 2010, LNCS 6684, 338–355.